\documentclass[a4paper,aps,prd,10pt,preprintnumbers,showpacs,onecolumn,superscriptaddress,nofootinbib,amsmath,amssymb]{revtex4-1}
\usepackage[dvips]{graphics}
\usepackage{cmap}
\usepackage[utf8]{inputenc}
\usepackage[T1]{fontenc}

\newcommand\diff{\mathrm{d}}

\begin{document}
\linespread{1.667}

\title{Quasinormal modes of the bumblebee black holes with a global monopole}
\author{Zainab Malik}\email{zainabmalik8115@gmail.com}
\affiliation{Institute of Applied Sciences and Intelligent Systems, H-15, Pakistan}
%

\begin{abstract}
We compute quasinormal modes of test fields around spherically symmetric black holes  with a global monopole in bumblebee gravity. The frequency of oscillation and the damping rate exhibit significant decreasing as the global monopole parameter is increased. An intriguing observation arises in the extreme limit, where the quasinormal modes manifest a form of universal behavior: the actual oscillation frequency remains unaltered despite variations in the Lorentz symmetry breaking (LSB) parameter. Our calculations are conducted through two distinct methods, both of which yield results that align remarkably well. Furthermore, we derive an analytical formula for quasinormal modes within the eikonal approximation and beyond it. In the limits of either vanishing deficit angle or bumblebee parameters, the compact and sufficiently accurate analytic expressions for quasinormal modes are obtained.
\end{abstract}

\maketitle

\section{Introduction}

Bumblebee Gravity is an unconventional theory proposing deviations from Einstein's General Relativity, involving the violation of Lorentz invariance and introducing additional fields, within the framework of effective field theory. This concept suggests that gravity might exhibit altered behavior under extreme conditions or at small scales, connecting it to the quest for a theory of quantum gravity. The theory's implications span cosmology, potentially explaining challenging observations, and it raises questions about dark matter, strong gravitational fields, and unresolved cosmological puzzles. The prospect of experimental or observational tests is a prominent feature of discussions surrounding "Bumblebee Gravity," reflecting the dynamic and exploratory nature of theoretical physics.  \cite{Bailey:2006fd,Bluhm:2004ep,Kostelecky:2010ze,Bluhm:2007bd,Ovgun:2018xys,Casana:2017jkc,Bertolami:2005bh,Li:2020dln,Ovgun:2018ran,Li:2020wvn}.

In the Bumblebee Gravity deviations from General Relativity suggest that black holes structure and surrounding accretion processes may differ from conventional models. The introduction of extra fields and violations of Lorentz invariance could lead to modified dynamics near black holes, impacting phenomena like accretion disk formation and radiation emission. Therefore, special attention was devoted to proper oscillation frequencies of black holes in this and other theories with broken Lorentz invariance - quasinormal modes \cite{Oliveira:2018oha,Kanzi:2021cbg,Oliveira:2021abg,Kanzi:2022vhp,Maluf:2022knd,Gogoi:2022wyv,Uniyal:2022xnq,Liu:2022dcn,Chen:2023cjd,Mangut:2023oxa,Filho:2023etf,Konoplya:2006rv,Konoplya:2006ar,Ding:2017gfw,Ding:2018mmh,Xiong:2021cth}.

Quasinormal modes (QNMs) of black holes are a fundamental aspect of their behavior, unveiling unique insights into the vibrational properties of black holes \cite{Kokkotas:1999bd,Nollert:1999ji,Konoplya:2011qq}.  In the realm of general relativity, QNMs represent the characteristic oscillations that arise when a perturbation disturbs the equilibrium of a black hole, leading to damped, exponentially decaying waveforms. These modes serve as a signature of black hole spacetime, revealing (together with the electromagnetic spectra \cite{Barack:2018yly,EventHorizonTelescope:2022xqj,Goddi:2016qax}) information about its mass, angular momentum, and charge. QNMs are pivotal in studying the stability and dynamics of black holes, as well as probing the nature of gravity and matter in extreme conditions. Analyzing QNMs facilitates the detection and interpretation of gravitational waves from black hole mergers, offering a powerful tool for testing theories of gravity and exploring the properties of black hole spacetime in various astrophysical scenarios.

One of the aspects we are interested here is black hole solution with a global monopole, as proposed by Vilenkin \cite{Vilenkin1981}, introduces a scenario where the presence of a topological defect affects the geometry of spacetime near a black hole. This solution, stemming from the interplay between gravitational and scalar fields, sheds light on the intriguing coupling between geometry and fundamental forces in certain theoretical frameworks.  While the quasinormal modes and grey-body factors of black holes in the Bumblebee Gravity  has been studied in a number of  works, no such studies was proposed for the case of the additional factor of the global monopole, except a recent study in \cite{Lin:2023foj} where the case of the scalar field was considered.

In this analysis, we will thoroughly examine the scenario characterized by asymptotic flatness. We accomplish this by investigating the propagation of scalar and neutrino fields across a range of parameter values, encompassing the quasi-extremal regime. By doing so, we corroborate and augment the findings outlined in a recent publication \cite{Lin:2023foj}. We ascertain quasinormal modes utilizing both the WKB method and time-domain integration, achieving a notable alignment with results from the Leaver method.

A notable observation is the pronounced decrease in the frequency of oscillation and damping rate as the global monopole parameter is augmented. In the limit of extreme black hole, the quasinormal modes exhibit a kind of uniform behavior: the actual oscillation frequency $Re (\omega)$ is unaffected by variations in the Lorentz symmetry breaking (LSB) parameter. Furthermore, we will derive an analytical formula for quasinormal modes within the eikonal approximation and at a few order beyond it. The latter analytical expression can serve as robust approximation for black holes with the explicit dependence of the frequencies upon the black hole parameters.

The structure of this paper is outlined as follows. In Section II, we deduce the wave-like equation and present the fundamental aspects of the methodologies utilized for identifying quasinormal modes. In Section III, we delve into a discussion regarding the quasinormal modes. In sec. IV and V the analytic formulas for quasinormal modes are derived using the expansion beyond the eikonal limit. Finally, we consolidate our findings within the Conclusions section.

\section{The wave-like equation and methods for study quasinormal modes}

\subsection{The metric}

We have two corrections to the Schwarzschild spacetime. The first correction is  produced by the local Lorenz Symmetry Breaking (LSB) due to the introduction of some vector field \cite{Casana:2017jkc} and characterized by the parameter $L$, which is, in other words, the bumblebee parameter. The other correction is induced by the global symmetry braking resulting to topological defects (see, for instance, \cite{Vilenkin:1981kz,Barriola:1989hx}). The latter is expressed by the constant  $k \eta ^{2}$ introduced in \cite{Gullu:2020qzu}, which can be interpreted as the deficit angle. 

Thus, the spherically symmetric black hole has the line element
\begin{equation}
    \label{sphansatz}
    \diff s^2=f(r)\diff t^2-\frac{\diff r^2}{h(r)}-r^2\diff\theta^2-r^2\sin^2\theta\diff\phi^2
\end{equation}
where $f(r)$ and $h(r)$ are functions of the radial coordinate $r$.
Here it is assumed that the bumblebee field is purely radial.
Under this supposition  the metric function describing a black hole has the form \cite{Gullu:2020qzu}
\begin{equation}\label{schlike}
    \begin{split}
        f(r)=   & 1-k \eta ^{2}-\frac{2M}{r}, \\
        h(r)=   & \frac{f(r)}{L+1},\\
    \end{split}
\end{equation}
where $L\equiv\xi b^2$ is the LSB parameter,
and $M$ is  the mass parameter which coincides with the Arnowitt-Deser-Misner in the Schwarzschild limit.
The event horizon is located  at $r_h=2M/(1-k \eta^2)$, independently of the value of the bumblebee parameter.

\subsection{Wave-like equation}

We will consider the test scalar field $\Phi$ which, thereby, is described by the covariant Klein-Gordon equation,
\begin{equation}\label{K-G}
    \frac{1}{\sqrt{-g}}\partial _{\mu }\left(g^{\mu\nu}\sqrt{-g}\partial_{\nu}\Phi \right)=0.
\end{equation}
Using the separation of variable,
\begin{equation}\label{separation1}
    \Phi=\frac{\psi_{l}(r)}{r}Y_{lm}(\theta ,\phi ) e^{-i\omega t},
\end{equation}
the dynamical equation for the scalar field can be reduced to the following wave-like equation:
\begin{equation}\label{radial2}
    \frac{\diff^2 \psi_{l}}{\diff x^2} +\left [  \omega ^{2}-U(x) \right ]\psi_{l}=0.
\end{equation}
Here, the effective potential is (see, for example, \cite{Abdalla:2006qj,Konoplya:2022iyn} for the deduction of the effective potential for the case of two metric functions)
\begin{equation}\label{potential}
    U(r)= f(r) \frac{\ell(\ell+1)}{r^2} + (s -1) \frac{(f(r) h(r))'}{2 r},
\end{equation}
where $s=0$ for the scalar field. Here $\ell$ is the multipole number and runs the following values: $\ell=0, 1, 2,...$ for $s=0$ and $\ell=1, 2, 3,..$ for $s=1$. Notice that when $s=1$ and the second term in the effective potential vanishes, the potential of the test electromagentic field is reproduced, which will be considered in the last two sections when deriving the analytical formulas beyond the eikonal approximation.
The tortoise coordinate $x$ is
\begin{equation}\label{toordinate}
    \dfrac{\diff x}{\diff r}=\frac{\sqrt{1+L} }{f(r)},
\end{equation}

Similarly, for the general covariant Dirac field we have
\begin{equation}\label{dirac}
            [\gamma^ae^\mu_a(\partial_\mu+\Gamma_\mu)]\Psi=0,
\end{equation}
  where $\gamma^a$ are the Dirac matrices, $e^\mu_a$ is the inverse of the tetrad $e^a_\mu$
  ($g_{\mu\nu} = \eta_{ab}e^a_\mu e^b_\nu$), $\eta_{ab}$ is the Minkowski metric.
  The spin connections $\Gamma_\mu$ are defined as follows:
\begin{equation}
		\Gamma_\mu=\frac{1}{8}[\gamma^a, \gamma^b]e^\nu_ae_{b\nu;\mu}.
\end{equation}
The above equation can be reduced to a couple of wave-like equations with the following effective potentials
\begin{equation}
  V_{\pm 1/2} = \frac{\sqrt{\tilde{f}}|\kappa|}{r^2}\Big(|\kappa|\sqrt{\tilde{f}} \pm \frac{r}{2}\frac{d\tilde{f}}{dr} \mp \tilde{f}\Big),
\end{equation}
where
$$ \tilde{f(r)} = f(r)/\sqrt{L+1}. $$
The number $\kappa= \ell+ \frac{1}{2}$ in the above equations are integer, while $\ell$ is half integer for fermions (see, for example, eq. (9) in \cite{Zhidenko:2003wq}).

\subsection{Higher order WKB method}

The main method we use here is the semi-analytic WKB method, applied for the first  time for finding quasinormal modes by B. Schutz and C. Will in \cite{Schutz:1985km}. The Schutz-Will formula was then extended to higher orders \cite{Iyer:1986np,Konoplya:2003ii,Matyjasek:2017psv} and made even more accurate when using the Padé approximants \cite{Matyjasek:2017psv}.
The general WKB formula has the form \cite{Konoplya:2019hlu},
\begin{equation}
\omega^2=V_0+A_2(K^2)+A_4(K^2)+A_6(K^2)+...
- i K\sqrt{-2V_2}\left(1+A_3(K^2)+A_5(K^2)+A_7(K^2)+...\right),
\end{equation}
where $K=n+1/2$ is half-integer. The corrections $A_k(K^2)$ of the order $k$ to the eikonal formula are polynomials of $K^2$ with rational coefficients. They depend on the values of higher derivatives of the effective potential $V(r)$ in its peak. With the procedure of Matyjasek and Opala \cite{Matyjasek:2017psv} using the Padé approximants the whole procedure is even more accurate. Here we will use the sixth order WKB method with $\tilde{m}=4$ and $5$, where the definition of $\tilde{m}$ and all details on this method can be found in \cite{Matyjasek:2017psv,Konoplya:2019hlu}. Higher order WKB method has been used for finding quasinormal modes and grey-body factors in great number of publications ( see for example \cite{Konoplya:2023ahd,Konoplya:2023ppx,Matyjasek:2021xfg,Guo:2023ivz,Chrysostomou:2022evl,Konoplya:2019xmn,Konoplya:2019ppy,Konoplya:2010vz,Matyjasek:2020bzc,Konoplya:2006gq} and references therein)  and comparison with accurate, convergent Leaver method  proved that the WKB formula is sufficiently accurate in the majority of cases \cite{Konoplya:2003ii}.

\subsection{Time-domain integration}

The integration in the time domain that we employ is founded on the wavelike equation expressed using the light-cone variables $u=t-r_*$ and $v=t+r_*$. We implement the discretization approach introduced by Gundlach-Price-Pullin \cite{Gundlach:1993tp} as follows:
\begin{equation}\label{Discretization}
\Psi\left(N\right)=\Psi\left(W\right)+\Psi\left(E\right)-\Psi\left(S\right) -\Delta^2V\left(S\right)\frac{\Psi\left(W\right)+\Psi\left(E\right)}{4}+{\cal O}\left(\Delta^4\right)\,,
\end{equation}
Here, the notation for the points is given by:
$N\equiv\left(u+\Delta,v+\Delta\right)$, $W\equiv\left(u+\Delta,v\right)$, $E\equiv\left(u,v+\Delta\right)$, and $S\equiv\left(u,v\right)$. The Gaussian initial data are imposed on the two null surfaces, specifically $u=u_0$ and $v=v_0$. The quasinormal frequencies can be extracted from the time-domain profiles reasonably accurately using the Prony method, as detailed in, for instance,~\cite{Konoplya:2011qq}. Although the time-domain integration method is indeed an accurate and convergent technique for generating time-domain profiles, the extraction of frequencies with high precision can be challenging. Generally, achieving a convergence with the exact results from the Leaver method, with a discrepancy of less than one percent, is attainable for the first and higher multipoles \cite{Bronnikov:2019sbx,Konoplya:2020jgt}.

\begin{table}
\begin{tabular}{|c|c|c|c|}
  \hline
    \hline
  $k \eta ^{2}$ & \text{6WKB} & \text{6WKB(Pade-5)} & \text{6WKB(Pade-6)}  \\
  \hline
  0. & \text{0.479592-0.078865 i} & \text{0.479593-0.078864 i} &   \text{0.479592-0.078864 i} \\
 0.1 & \text{0.408788-0.063857 i} & \text{0.408789-0.063856 i}   & \text{0.408788-0.063856 i} \\
 0.2 & \text{0.342003-0.050437 i} & \text{0.342003-0.050436 i}   & \text{0.342003-0.050436 i} \\
 0.3 & \text{0.279447-0.038601 i} & \text{0.279447-0.038601 i}   & \text{0.279447-0.038601 i} \\
 0.4 & \text{0.221379-0.028349 i} & \text{0.221379-0.028349 i}   & \text{0.221379-0.028349 i} \\
 0.5 & \text{0.168119-0.019680 i} & \text{0.168119-0.019680 i}   & \text{0.168119-0.019680 i} \\
 0.6 & \text{0.1200890-0.0125902 i} & \text{0.1200890-0.0125902   i} & \text{0.1200890-0.0125902 i} \\
 0.7 & \text{0.0778652-0.0070793 i} & \text{0.0778652-0.0070793   i} & \text{0.0778652-0.0070793 i} \\
 0.8 & \text{0.0423110-0.0031451 i} & \text{0.0423110-0.0031451   i} & \text{0.0423110-0.0031451 i} \\
 0.9 & \text{0.0149332-0.0007860 i} & \text{0.0149332-0.0007860   i} & \text{0.0149332-0.0007860 i} \\
 0.94 & \text{0.00693547-0.00028291 i} &   \text{0.00693547-0.00028291 i} & \text{0.00693547-0.00028291   i} \\
 0.98 & \text{0.001333800-0.000031429 i} &   \text{0.001333800-0.000031429 i} &   \text{0.001333800-0.000031429 i}\\
  \hline
    \hline
\end{tabular}
\caption{Fundamental quasinormal mode for the scalar field for $\ell=2$, $L=0.5$ and various values of $k \eta ^{2}$. }
\end{table}

\begin{table}
\begin{tabular}{|c|c|c|c|}
  \hline
    \hline
  $k \eta ^{2}$ & \text{6WKB} & \text{6WKB(Pade-5)} & \text{6WKB(Pade-6)}  \\
  \hline
  0. & \text{0.477556-0.068236 i} & \text{0.477556-0.068236 i} &   \text{0.477556-0.068236 i} \\
 0.1 & \text{0.407222-0.055256 i} & \text{0.407222-0.055255 i}   & \text{0.407222-0.055255 i} \\
 0.2 & \text{0.340835-0.043647 i} & \text{0.340835-0.043646 i}   & \text{0.340835-0.043646 i} \\
 0.3 & \text{0.278610-0.033408 i} & \text{0.278610-0.033407 i}   & \text{0.278610-0.033407 i} \\
 0.4 & \text{0.220808-0.024537 i} & \text{0.220808-0.024537 i}   & \text{0.220808-0.024537 i} \\
 0.5 & \text{0.167757-0.017035 i} & \text{0.167757-0.017035 i}   & \text{0.167757-0.017035 i} \\
 0.6 & \text{0.1198814-0.0108993 i} & \text{0.1198814-0.0108993   i} & \text{0.1198814-0.0108993 i} \\
 0.7 & \text{0.0777640-0.0061291 i} & \text{0.0777640-0.0061291   i} & \text{0.0777640-0.0061291 i} \\
 0.8 & \text{0.0422742-0.0027232 i} & \text{0.0422742-0.0027232   i} & \text{0.0422742-0.0027232 i} \\
 0.9 & \text{0.0149267-0.0006806 i} & \text{0.0149267-0.0006806   i} & \text{0.0149267-0.0006806 i} \\
 0.94 & \text{0.00693365-0.00024499 i} &   \text{0.00693365-0.00024499 i} & \text{0.00693365-0.00024499   i} \\
 0.98 & \text{0.001333683-0.000027218 i} &   \text{0.001333683-0.000027218 i} &   \text{0.001333683-0.000027218 i}\\
  \hline
    \hline
\end{tabular}
\caption{Fundamental quasinormal mode for the scalar field for $\ell=2$, $L=1$ and various values of $k \eta ^{2}$. }
\end{table}

\begin{table}
\begin{tabular}{|c|c|c|c|}
  \hline
    \hline
  $k \eta ^{2}$ & \text{6WKB} & \text{6WKB(Pade-5)} & \text{6WKB(Pade-6)}  \\
  \hline
 0. & \text{0.475513-0.055662 i} & \text{0.475513-0.055662 i} &   \text{0.475513-0.055662 i} \\
 0.1 & \text{0.405650-0.045078 i} & \text{0.405650-0.045078 i}   & \text{0.405650-0.045078 i} \\
 0.2 & \text{0.339663-0.035610 i} & \text{0.339663-0.035610 i}   & \text{0.339663-0.035610 i} \\
 0.3 & \text{0.277770-0.027259 i} & \text{0.277770-0.027259 i}   & \text{0.277770-0.027259 i} \\
 0.4 & \text{0.220236-0.020023 i} & \text{0.220236-0.020023 i}   & \text{0.220236-0.020023 i} \\
 0.5 & \text{0.167394-0.013902 i} & \text{0.167394-0.013902 i}   & \text{0.167394-0.013902 i} \\
 0.6 & \text{0.1196735-0.0088958 i} & \text{0.1196735-0.0088958   i} & \text{0.1196735-0.0088958 i} \\
 0.7 & \text{0.0776627-0.0050029 i} & \text{0.0776627-0.0050029  i} & \text{0.0776627-0.0050029 i} \\
 0.8 & \text{0.0422374-0.0022231 i} & \text{0.0422374-0.0022231   i} & \text{0.0422374-0.0022231 i} \\
 0.9 & \text{0.0149202-0.0005557 i} & \text{0.0149202-0.0005557   i} & \text{0.0149202-0.0005557 i} \\
  0.94 & \text{0.00693184-0.00020002 i} &   \text{0.00693184-0.00020002 i} & \text{0.00693184-0.00020002   i} \\
 0.98 & \text{0.001333567-0.000022223 i} &   \text{0.001333567-0.000022223 i} &   \text{0.001333567-0.000022223 i}\\
  \hline
    \hline
\end{tabular}
\caption{Fundamental quasinormal mode for the scalar field for $\ell=2$, $L=2$ and various values of $k \eta ^{2}$. }
\end{table}

\section{Quasinormal modes}

\begin{table}
\begin{tabular}{|c|c|c|c|}
  \hline
    \hline
  $k \eta ^{2}$ & \text{6WKB} & \text{6WKB(Pade-5)} & \text{6WKB(Pade-6)}  \\
  \hline
 0. & \text{0.282706-0.068607 i} & \text{0.282711-0.068593 i} &   \text{0.282710-0.068592 i} \\
 0.1 & \text{0.240492-0.055528 i} & \text{0.240495-0.055519 i}   & \text{0.240494-0.055518 i} \\
 0.2 & \text{0.200799-0.043839 i} & \text{0.200801-0.043834 i}   & \text{0.200800-0.043833 i} \\
 0.3 & \text{0.163739-0.033537 i} & \text{0.163740-0.033534 i}   & \text{0.163740-0.033534 i} \\
 0.4 & \text{0.1294496-0.0246193 i} & \text{0.1294500-0.0246180   i} & \text{0.1294498-0.0246179 i} \\
 0.5 & \text{0.0981035-0.0170827 i} & \text{0.0981037-0.0170822   i} & \text{0.0981036-0.0170821 i} \\
 0.6 & \text{0.0699301-0.0109239 i} & \text{0.0699301-0.0109237   i} & \text{0.0699301-0.0109237 i} \\
 0.7 & \text{0.0452468-0.0061395 i} & \text{0.0452469-0.0061395   i} & \text{0.0452469-0.0061395 i} \\
 0.8 & \text{0.0245342-0.0027264 i} & \text{0.0245342-0.0027264   i} & \text{0.0245342-0.0027264 i} \\
 0.9 & \text{0.00864045-0.00068101 i} &   \text{0.00864045-0.00068101 i} & \text{0.00864045-0.00068101   i} \\
  0.94 & \text{0.00400944-0.00024508 i} &   \text{0.00400944-0.00024508 i} & \text{0.00400944-0.00024508   i} \\
 0.98 & \text{0.000770406-0.000027221 i} &  \text{0.000770406-0.000027221 i} &   \text{0.000770406-0.000027221 i}\\
  \hline
    \hline
\end{tabular}
\caption{Fundamental quasinormal mode for the scalar field $\ell=1$, $L=1$ and various values of $k \eta ^{2}$. }
\end{table}

\begin{table}
\begin{tabular}{|c|c|c|c|}
  \hline
    \hline
  $k \eta ^{2}$ & \text{6WKB} & \text{6WKB(Pade-5)} & \text{6WKB(Pade-6)}  \\
  \hline
 0. & \text{0.186091-0.079016 i} & \text{0.186060-0.078971 i} &   \text{0.186043-0.078977 i} \\
 0.1 & \text{0.159413-0.063965 i} & \text{0.159395-0.063933 i}   & \text{0.159385-0.063936 i} \\
 0.2 & \text{0.134041-0.050510 i} & \text{0.134031-0.050488 i}   & \text{0.134025-0.050490 i} \\
 0.3 & \text{0.1100775-0.0386487 i} & \text{0.1100726-0.0386348   i} & \text{0.1100694-0.0386360 i} \\
 0.4 & \text{0.0876462-0.0283780 i} & \text{0.0876441-0.0283700   i} & \text{0.0876426-0.0283707 i} \\
 0.5 & \text{0.0668992-0.0196952 i} & \text{0.0668985-0.0196913   i} & \text{0.0668979-0.0196916 i} \\
 0.6 & \text{0.0480309-0.0125976 i} & \text{0.0480307-0.0125960   i} & \text{0.0480305-0.0125961 i} \\
 0.7 & \text{0.0313027-0.0070822 i} & \text{0.0313026-0.0070817   i} & \text{0.0313026-0.0070817 i} \\
 0.8 & \text{0.0170968-0.0031459 i} & \text{0.0170968-0.0031458   i} & \text{0.0170968-0.0031458 i} \\
 0.9 & \text{0.00606518-0.00078607 i} &   \text{0.00606518-0.00078606 i} & \text{0.00606518-0.00078606   i}\\
 0.94 & \text{0.00282267-0.00028293 i} &   \text{0.00282267-0.00028293 i} & \text{0.00282267-0.00028293   i} \\
 0.98 & \text{0.000543962-0.000031430 i} &   \text{0.000543962-0.000031430 i} &   \text{0.000543962-0.000031430 i}\\
  \hline
    \hline
\end{tabular}
\caption{Fundamental quasinormal mode for the Dirac field for $\ell=1$, $L=0.5$ and various values of $k \eta ^{2}$. }
\end{table}

\begin{table}
\begin{tabular}{|c|c|c|c|}
  \hline
    \hline
  $k \eta ^{2}$ & \text{6WKB} & \text{6WKB(Pade-5)} & \text{6WKB(Pade-6)}  \\
  \hline
0. & \text{0.187640-0.068328 i} & \text{0.187629-0.068301 i} &   \text{0.187623-0.068303 i} \\
 0.1 & \text{0.160612-0.055321 i} & \text{0.160606-0.055302 i}   & \text{0.160602-0.055304 i} \\
 0.2 & \text{0.134940-0.043691 i} & \text{0.134937-0.043678 i}   & \text{0.134935-0.043679 i} \\
 0.3 & \text{0.1107255-0.0334358 i} & \text{0.1107240-0.0334284  i} & \text{0.1107228-0.0334290 i} \\
 0.4 & \text{0.0880896-0.0245543 i} & \text{0.0880890-0.0245502   i} & \text{0.0880885-0.0245505 i} \\
 0.5 & \text{0.0671819-0.0170442 i} & \text{0.0671818-0.0170423   i} & \text{0.0671816-0.0170424 i} \\
 0.6 & \text{0.0481936-0.0109037 i} & \text{0.0481936-0.0109029   i} & \text{0.0481935-0.0109030 i} \\
 0.7 & \text{0.0313823-0.0061308 i} & \text{0.0313823-0.0061306   i} & \text{0.0313823-0.0061306 i} \\
 0.8 & \text{0.0171259-0.0027237 i} & \text{0.0171259-0.0027237   i} & \text{0.0171259-0.0027237 i} \\
 0.9 & \text{0.00607033-0.00068067 i} &   \text{0.00607033-0.00068067 i} & \text{0.00607033-0.00068067 i}\\
  0.94 & \text{0.00282411-0.00024500 i} &   \text{0.00282411-0.00024500 i} & \text{0.00282411-0.00024500   i} \\
 0.98 & \text{0.000544054-0.000027219 i} &  \text{0.000544054-0.000027219 i} &   \text{0.000544054-0.000027219 i}\\
  \hline
    \hline
\end{tabular}
\caption{Fundamental quasinormal mode for the Dirac field for $\ell=1$, $L=1$ and various values of $k \eta ^{2}$. }
\end{table}

Our primary objective is to establish a comparative analysis between the numerical data obtained in our work and those recently published in \cite{Lin:2023foj}. Regrettably, \cite{Lin:2023foj} does not present numerical data in tabular form, with the exception of a sample illustration involving specific parameter values: $L=0.5$, $\ell=2$, $M=1$, and $k  \eta ^{2}=0.2$. The quasinormal modes were computed in \cite{Lin:2023foj} utilizing both the time-domain integration and the continued fraction method (which seemingly corresponds to the Leaver method \cite{Leaver:1985ax}). The following values were obtained for the aforementioned parameters:
$$\omega = 0.342031 - 0.0504228 i \quad (time-domain~LJZ),$$
$$\omega = 0.342003 - 0.0504359 i \quad (continued~fraction~LJZ),$$
where "LJZ" signifies the Lin-Jiang-Zhai data of \cite{Lin:2023foj}.

This value of $\omega$ aligns with the depiction in figure 3 of \cite{Lin:2023foj}, hence we will use it for comparison.
Ordinarily, the precision of the time-domain integration method is limited the period of quasinormal oscillations is short. The exact determination of this period becomes challenging due to the interplay between the initial outburst and the asymptotic tails. Therefore, it was rather surprising to observe such a remarkable concurrence between the time-domain integration approach and the Leaver method, which is renowned for its accuracy.

For the same values of the parameters we obtained the following values of the quasinormal modes:
$$\omega = 0.342468 - 0.0504003 i  \quad (time-domain),$$
$$\omega = 0.342003-0.050437 i \quad (WKB~6th~order),$$
$$\omega = 0.342003-0.050436 i \quad (WKB~6th~order~Pade).$$

As anticipated, the time-domain method demonstrates a remarkably close agreement with the 6th order WKB formula when employing Pade approximants. The relative error remains a minute fraction, well within the confines of one percent. The accuracy of the WKB approach diminishes when Pade approximants are excluded. Remarkably, our WKB data, complemented by Pade approximants, exhibit an impressive alignment with the meticulously accurate data obtained through the continued fraction method in \cite{Lin:2023foj}.

Upon examining tables I-VI, it becomes evident that the foremost parameter influencing the quasinormal frequencies is the global monopole parameter $k \eta ^{2}$. This parameter significantly suppresses both the real and imaginary components of the modes. The Lorentz symmetry breaking (LSB) parameter exerts a milder influence but also results in a decrease in the quasinormal modes. Of particular interest is the observation that, for a substantial impact of the global monopole (leading to near-extreme states, e.g., $k \eta ^{2} =0.98$), the real component of the quasinormal modes becomes independent of the other parameter $L$.

Quasinormal modes can be found in the analytical approximate form in the regime of large values of the multipole moment $\ell$, which corresponds to high frequencies of oscillations. Indeed, using the expression for the peak of the effective potential in the eikonal regime,
$$ r_{max} = \frac{3 M} {1-k \eta ^{2}},$$
and substituting it into the first order WKB formula
\begin{equation}
 i \frac{(\omega^2 - V_{\max})}{\sqrt{-2 V''_{\max}}} =0,
\end{equation}
we find an expression for the quasinormal mode at high $l$:
\begin{equation}
\omega = -\frac{\left(\ell+\frac{1}{2}\right) \sqrt{L+1} \left(\eta
   ^2 k -1\right)^3+i
   (L+1)
   \left(n+\frac{1}{2}\right)
   \sqrt{-\frac{\left(\eta ^2
   k
   -1\right)^7}{(L+1)^{3/2}}}}
   {3 \sqrt{3} (L+1) M
   \sqrt{-\frac{\left(\eta ^2
   k
   -1\right)^3}{\sqrt{L+1}}}}
\end{equation}
When $L=\eta ^2 k=0$ we reproduce the Schwarzschild limit,
$$ \omega = \frac{\ell-i
   \left(n+\frac{1}{2}\right)+
   \frac{1}{2}}{3 \sqrt{3} M}.$$

The aforementioned eikonal formula holds particular interest due to its alignment with the characteristics of null geodesics pertinent to test fields, encompassing the Lyapunov exponent and frequency at the unstable circular orbit \cite{Cardoso:2008bp}. Consequently, this formula also maintains a connection with the diameter of the shadow cast by a black hole \cite{Jusufi:2019ltj,Jafarzade:2020ova}. While \cite{Konoplya:2017wot,Konoplya:2022gjp} demonstrates that the correspondence is typically upheld, there are instances where it can be disrupted, such as for certain non-minimally coupled fields, or it might remain incomplete for the asymptotically de Sitter scenario. However, one can readily verify through calculations of the pertinent geodesic parameters that the correspondence indeed holds true within our specific case.

The correspondence between null geodesics and eikonal quasinormal modes states that
\begin{equation}\label{QNM}
\omega_n = \Omega \ell-i (n+1/2)|\lambda|, \quad \ell \gg n.
\end{equation}
Here $\Omega$ is the angular velocity at the unstable null geodesics, and $\lambda$ is the Lyapunov exponent.
The radius of the unstable circular null geodesic $r_inst$ is at $V'(r)=0$. Therefore, following  \cite{Cardoso:2008bp} we have
\begin{equation}\label{radius}
2 f(r_{inst}) = r_{inst} f'(r)_{r=r_{inst}},
\end{equation}
while the angular velocity is
\begin{equation}\label{omega}
\Omega=\frac{d \phi} {d t} = \frac{f(r_{inst})^{1/2}}{r_{inst}}.
\end{equation}
Thus, we conclude that the correspondence is confirmed for the Dirac field as well.

\section{Analytic formula in the limit of vanishing bumblebee parameter}

Using the expansion in powers of $1/\ell$ as in \cite{Konoplya:2023moy},
we can find compact and sufficiently accurate analytic approximations for quasinormal modes in the limit when the bumblebee parameter is zero, that is, for the black hole with a deficit angle. This approach has been recently applied by the author to find sufficient accurate and compact analytic epxressions for quasinormal modes of dilaton
\cite{Malik:2024sxv} and Reissner-Nordstrom-like black holes \cite{Malik:2024voy}.

For the scalar field we find  the position of the peak of the effective potential in the form of the following series expansion,
\begin{equation}\label{rmax-scalar}
r_{\max } = 3 M-\frac{M}{3 \kappa ^2}+3 M q+3 M q^2+3 M q^3+\mathcal{O}\left(q^4,\frac{1}{\kappa ^4}\right).
\end{equation}
Here, for compactness, we use a new designation $q = k \eta ^{2}$
Then, using the 6th order WKB formula, we obtain the expression for the quasinormal modes
\begin{equation}\label{eikonal-scalar}
\begin{array}{rcl}
\omega  &=& \displaystyle-\frac{i K \left(940 K^2+313\right)}{46656 \sqrt{3} M \kappa ^2}+\frac{29-60 K^2}{1296 \sqrt{3} M \kappa }+\frac{\kappa }{3 \sqrt{3} M}-\frac{i K}{3 \sqrt{3} M}\\
&&\displaystyle+q \left(\frac{i K \left(940 K^2+313\right)}{15552 \sqrt{3} M \kappa ^2}+\frac{300 K^2-253}{2592 \sqrt{3} M \kappa }-\frac{\kappa }{2 \sqrt{3} M}+\frac{2 i K}{3 \sqrt{3} M}\right)\\
&&\displaystyle+q^2 \left(-\frac{i K \left(940 K^2+313\right)}{15552 \sqrt{3} M \kappa ^2}+\frac{361-300 K^2}{3456 \sqrt{3} M \kappa }+\frac{\kappa }{8 \sqrt{3} M}-\frac{i K}{3 \sqrt{3} M}\right)\\
&&\displaystyle+q^3 \left(\frac{i K \left(940 K^2+313\right)}{46656 \sqrt{3} M \kappa ^2}+\frac{300 K^2-469}{20736 \sqrt{3} M \kappa }+\frac{\kappa }{48 \sqrt{3} M}\right)+\mathcal{O}\left(q^4,\frac{1}{\kappa ^3}\right)
\end{array}
\end{equation}
where $\kappa\equiv\ell+1/2$, $K\equiv n+1/2$.

In the same way, the potential peak expansion for the electromagnetic field has the form
\begin{equation}\label{rmax-electromagnetic}
\begin{array}{rcl}
r_{\max } &=& \displaystyle\frac{11 M}{16 \sqrt{3} \kappa ^3}-\frac{\sqrt{3} M}{2 \kappa }+3 M\\
&&\displaystyle+q \left(-\frac{11 M}{32 \sqrt{3} \kappa ^3}-\frac{\sqrt{3} M}{4 \kappa }+3 M\right)\\
&&\displaystyle+q^2 \left(-\frac{11 M}{128 \sqrt{3} \kappa ^3}-\frac{3 \sqrt{3} M}{16 \kappa }+3 M\right)\\
&&\displaystyle+q^3 \left(-\frac{11 M}{256 \sqrt{3} \kappa ^3}-\frac{5 \sqrt{3} M}{32 \kappa }+3 M\right)+\mathcal{O}\left(q^4,\frac{1}{\kappa ^4}\right)
\end{array}
\end{equation}
and the consequent formula for the quasinormal modes is
\begin{equation}\label{eikonal-electromagnetic}
\begin{array}{rcl}
\omega  &=& \displaystyle\frac{i K \left(119-940 K^2\right)}{46656 \sqrt{3} M \kappa ^2}-\frac{60 K^2+7}{1296 \sqrt{3} M \kappa }+\frac{\kappa }{3 \sqrt{3} M}-\frac{i K}{3 \sqrt{3} M}\\
&&\displaystyle+q \left(\frac{i K \left(940 K^2-119\right)}{15552 \sqrt{3} M \kappa ^2}+\frac{5 \left(60 K^2+7\right)}{2592 \sqrt{3} M \kappa }-\frac{\kappa }{2 \sqrt{3} M}+\frac{2 i K}{3 \sqrt{3} M}\right)\\
&&\displaystyle+q^2 \left(\frac{i K \left(119-940 K^2\right)}{15552 \sqrt{3} M \kappa ^2}-\frac{5 \left(60 K^2+7\right)}{3456 \sqrt{3} M \kappa }+\frac{\kappa }{8 \sqrt{3} M}-\frac{i K}{3 \sqrt{3} M}\right)\\
&&\displaystyle+q^3 \left(\frac{i K \left(940 K^2-119\right)}{46656 \sqrt{3} M \kappa ^2}+\frac{5 \left(60 K^2+7\right)}{20736 \sqrt{3} M \kappa }+\frac{\kappa }{48 \sqrt{3} M}\right)+\mathcal{O}\left(q^4,\frac{1}{\kappa ^3}\right)
\end{array}
\end{equation}

Finally, for the Dirac field,  the position of the maximum of the potential is
\begin{equation}\label{rmax-Dirac}
\begin{array}{rcl}
r_{\max } &=& \displaystyle\frac{11 M}{16 \sqrt{3} \kappa ^3}-\frac{\sqrt{3} M}{2 \kappa }+3 M\\
&&\displaystyle+q \left(-\frac{11 M}{32 \sqrt{3} \kappa ^3}-\frac{\sqrt{3} M}{4 \kappa }+3 M\right)\\
&&\displaystyle+q^2 \left(-\frac{11 M}{128 \sqrt{3} \kappa ^3}-\frac{3 \sqrt{3} M}{16 \kappa }+3 M\right)\\
&&\displaystyle+q^3 \left(-\frac{11 M}{256 \sqrt{3} \kappa ^3}-\frac{5 \sqrt{3} M}{32 \kappa }+3 M\right)+\mathcal{O}\left(q^4,\frac{1}{\kappa ^4}\right)
\end{array}
\end{equation}
and the quasinormal modes are
\begin{equation}\label{eikonal-Dirac}
\begin{array}{rcl}
\omega  &=& \displaystyle\frac{i K \left(119-940 K^2\right)}{46656 \sqrt{3} M \kappa ^2}-\frac{60 K^2+7}{1296 \sqrt{3} M \kappa }+\frac{\kappa }{3 \sqrt{3} M}-\frac{i K}{3 \sqrt{3} M}\\
&&\displaystyle+q \left(\frac{i K \left(940 K^2-119\right)}{15552 \sqrt{3} M \kappa ^2}+\frac{5 \left(60 K^2+7\right)}{2592 \sqrt{3} M \kappa }-\frac{\kappa }{2 \sqrt{3} M}+\frac{2 i K}{3 \sqrt{3} M}\right)\\
&&\displaystyle+q^2 \left(\frac{i K \left(119-940 K^2\right)}{15552 \sqrt{3} M \kappa ^2}-\frac{5 \left(60 K^2+7\right)}{3456 \sqrt{3} M \kappa }+\frac{\kappa }{8 \sqrt{3} M}-\frac{i K}{3 \sqrt{3} M}\right)\\
&&\displaystyle+q^3 \left(\frac{i K \left(940 K^2-119\right)}{46656 \sqrt{3} M \kappa ^2}+\frac{5 \left(60 K^2+7\right)}{20736 \sqrt{3} M \kappa }+\frac{\kappa }{48 \sqrt{3} M}\right)+\mathcal{O}\left(q^4,\frac{1}{\kappa ^3}\right).
\end{array}
\end{equation}


\begin{table}
\begin{tabular}{ c c c c }
\hline
\hline
q & WKB6 & analytic & difference \\
\hline
$0$ & $0.2929-0.0978 i$ & $0.2928-0.0977 i$ & $0.03\%$\\
$0.1$ & $0.2484-0.0791 i$ & $0.2483-0.0790 i$ & $0.03\%$\\
$0.2$ & $0.20670-0.06239 i$ & $0.207-0.0624 i$ & $0.03\%$\\
$0.3$ & $0.16798-0.04769 i$ & $0.168-0.048 i$ & $0.05\%$\\
$0.4$ & $0.13235-0.03498 i$ & $0.1322-0.0350 i$ & $0.141\%$\\
$0.5$ & $0.09995-0.02426 i$ & $0.099-0.0242 i$ & $0.46\%$\\
$0.6$ & $0.070993-0.015499 i$ & $0.070-0.015 i$ & $1.48\%$\\
\hline
\hline
\end{tabular}
\caption{Quasinormal modes of the $\ell=1$ test scalar field for the black hole with a deficit angle calculated using the 6th order WKB formula and using the approximate analytic eikonal formula. The deviation is given in per cents.}
\end{table}

\begin{table}
\begin{tabular}{ c c c c }
\hline
\hline
q & WKB6 & analytic & difference \\
\hline
$0$ & $0.1826-0.0949 i$ & $0.1826-0.0969 i$ & $0.98\%$\\
$0.1$ & $0.1570-0.0770 i$ & $0.1568-0.0785 i$ & $0.86\%$\\
$0.2$ & $0.1323-0.0609 i$ & $0.132-0.0620 i$ & $0.72\%$\\
$0.3$ & $0.1089-0.0468 i$ & $0.109-0.047 i$ & $0.58\%$\\
$0.4$ & $0.0869-0.0345 i$ & $0.0866-0.0348 i$ & $0.50\%$\\
$0.5$ & $0.06640-0.02400 i$ & $0.066-0.0241 i$ & $0.75\%$\\
$0.6$ & $0.04773-0.01539 i$ & $0.047-0.015 i$ & $1.92\%$\\
\hline
\hline
\end{tabular}
\caption{Quasinormal modes of the $\ell=1/2$ Dirac field for the black hole with a deficit angle calculated using the 6th order WKB formula and using the approximate analytic eikonal formula. The deviation is given in per cents.}
\end{table}

Comparison of the quasinromal frequencies calculated via the obtained analytic formulas are shown in tables VII-VIII and comparison with the 6th order WKB formula shows very good agreement for all $\ell$, except the lowest $\ell =0$ for the scalar field perturbations.

\section{Analytic formula in the limit of vanishing deficit angle}

\begin{table}
\begin{tabular}{ c c c c }
\hline
\hline
L & WKB6 & analytic & difference \\
\hline
$0$ & $0.2929-0.0978 i$ & $0.2928-0.0977 i$ & $0.03\%$\\
$0.1$ & $0.2775-0.0888 i$ & $0.277-0.0887 i$ & $0.03\%$\\
$0.2$ & $0.2643-0.0813 i$ & $0.2641-0.0811 i$ & $0.11\%$\\
$0.3$ & $0.2528-0.0749 i$ & $0.252-0.074 i$ & $0.41\%$\\
$0.4$ & $0.24266-0.06954 i$ & $0.240-0.0676 i$ & $1.20\%$\\
$0.5$ & $0.23363-0.06485 i$ & $0.2284-0.0605 i$ & $2.81\%$\\
$0.6$ & $0.22554-0.06076 i$ & $0.215-0.052 i$ & $5.65\%$\\
\hline
\hline
\end{tabular}
\caption{Quasinormal modes of the $\ell=1$ test scalar field ($s=0$) for the bumblebee black hole with zero deficit angle calculated using the 6th order WKB formula and using the approximate analytic eikonal formula. The deviation is given in per cents.}
\end{table}

\begin{table}
\begin{tabular}{ c c c c }
\hline
\hline
L & WKB6 & analytic & difference \\
\hline
$0$ & $0.2824-0.0961 i$ & $0.2821-0.0965 i$ & $0.18\%$\\
$0.1$ & $0.2698-0.0875 i$ & $0.270-0.0877 i$ & $0.13\%$\\
$0.2$ & $0.2587-0.0802 i$ & $0.2585-0.0803 i$ & $0.1\%$\\
$0.3$ & $0.2489-0.0741 i$ & $0.248-0.074 i$ & $0.28\%$\\
$0.4$ & $0.2401-0.0688 i$ & $0.239-0.0671 i$ & $0.86\%$\\
$0.5$ & $0.2322-0.0642 i$ & $0.2294-0.0602 i$ & $2.04\%$\\
$0.6$ & $0.2250-0.0602 i$ & $0.220-0.052 i$ & $4.11\%$\\
\hline
\hline
\end{tabular}
\caption{Quasinormal modes of the $\ell=1$ electromagnetic field ($s=1$) for the bumblebee black hole with zero deficit angle calculated using the 6th order WKB formula and using the approximate analytic eikonal formula. The deviation is given in per cents.}
\end{table}

\begin{table}
\begin{tabular}{ c c c c }
\hline
\hline
L & WKB6 & analytic & difference \\
\hline
$0$ & $0.1826-0.0949 i$ & $0.1826-0.0969 i$ & $0.98\%$\\
$0.1$ & $0.1752-0.0864 i$ & $0.175-0.0881 i$ & $0.86\%$\\
$0.2$ & $0.1685-0.0793 i$ & $0.1682-0.0806 i$ & $0.70\%$\\
$0.3$ & $0.1625-0.0733 i$ & $0.162-0.074 i$ & $0.41\%$\\
$0.4$ & $0.1571-0.0681 i$ & $0.1562-0.0673 i$ & $0.72\%$\\
$0.5$ & $0.1521-0.0637 i$ & $0.1505-0.0603 i$ & $2.26\%$\\
$0.6$ & $0.1476-0.0598 i$ & $0.145-0.052 i$ & $4.99\%$\\
\hline
\hline
\end{tabular}
\caption{Quasinormal modes of the $\ell=1/2$ Dirac field for the bumblebee black hole with zero deficit angle calculated using the 6th order WKB formula and using the approximate analytic eikonal formula. The deviation is given in per cents.}
\end{table}

In the same way, a compact analytic expression can be obtained in the limit of zero deficit angle.
Expanding in terms of the inverse multipole number \cite{Konoplya:2023moy}, we find the position of a maximum of the scalar field effective potential,
\begin{equation}\label{rmax-scalar}
r_{\max } = 3 M-\frac{M}{3 \kappa ^2}+\frac{M L}{3 \kappa ^2}-\frac{M L^2}{3 \kappa ^2}+\frac{M L^3}{3 \kappa ^2}+\mathcal{O}\left(L^4,\frac{1}{\kappa ^4}\right)
\end{equation}
and, using the WKB formula, the expression for scalar field's quasinormal modes
\begin{equation}\label{eikonal-scalar}
\begin{array}{rcl}
\omega  &=& \displaystyle-\frac{i K \left(940 K^2+313\right)}{46656 \sqrt{3} M \kappa ^2}+\frac{29-60 K^2}{1296 \sqrt{3} M \kappa }+\frac{\kappa }{3 \sqrt{3} M}-\frac{i K}{3 \sqrt{3} M}\\
&&\displaystyle+L \left(\frac{i K \left(940 K^2+313\right)}{23328 \sqrt{3} M \kappa ^2}+\frac{5 \left(12 K^2-13\right)}{864 \sqrt{3} M \kappa }-\frac{\kappa }{6 \sqrt{3} M}+\frac{i K}{3 \sqrt{3} M}\right)\\
&&\displaystyle+L^2 \left(-\frac{i K \left(940 K^2+313\right)}{15552 \sqrt{3} M \kappa ^2}+\frac{361-300 K^2}{3456 \sqrt{3} M \kappa }+\frac{\kappa }{8 \sqrt{3} M}-\frac{i K}{3 \sqrt{3} M}\right)\\
&&\displaystyle+L^3 \left(\frac{i K \left(940 K^2+313\right)}{11664 \sqrt{3} M \kappa ^2}+\frac{5 \left(420 K^2-527\right)}{20736 \sqrt{3} M \kappa }-\frac{5 \kappa }{48 \sqrt{3} M}+\frac{i K}{3 \sqrt{3} M}\right)+\mathcal{O}\left(L^4,\frac{1}{\kappa ^3}\right).
\end{array}
\end{equation}
Here we have $\kappa\equiv\ell+1/2$, $K\equiv n+1/2$.

The potential peak of the electromagnetic field is situated at
\begin{equation}\label{rmax-electromagnetic}
\begin{array}{rcl}
r_{\max } &=& \displaystyle\frac{11 M}{16 \sqrt{3} \kappa ^3}-\frac{\sqrt{3} M}{2 \kappa }+3 M\\
&&\displaystyle+L \left(\frac{\sqrt{3} M}{4 \kappa }-\frac{11 \sqrt{3} M}{32 \kappa ^3}\right)\\
&&\displaystyle+L^2 \left(\frac{55 \sqrt{3} M}{128 \kappa ^3}-\frac{3 \sqrt{3} M}{16 \kappa }\right)\\
&&\displaystyle+L^3 \left(\frac{5 \sqrt{3} M}{32 \kappa }-\frac{385 M}{256 \sqrt{3} \kappa ^3}\right)+\mathcal{O}\left(L^4,\frac{1}{\kappa ^4}\right),
\end{array}
\end{equation}
and the formula for the quasinormal modes has the following form
\begin{equation}\label{eikonal-electromagnetic}
\begin{array}{rcl}
\omega  &=& \displaystyle\frac{i K \left(119-940 K^2\right)}{46656 \sqrt{3} M \kappa ^2}-\frac{60 K^2+7}{1296 \sqrt{3} M \kappa }+\frac{\kappa }{3 \sqrt{3} M}-\frac{i K}{3 \sqrt{3} M}\\
&&\displaystyle+L \left(\frac{i K \left(940 K^2-119\right)}{23328 \sqrt{3} M \kappa ^2}+\frac{60 K^2+7}{864 \sqrt{3} M \kappa }-\frac{\kappa }{6 \sqrt{3} M}+\frac{i K}{3 \sqrt{3} M}\right)\\
&&\displaystyle+L^2 \left(\frac{i K \left(119-940 K^2\right)}{15552 \sqrt{3} M \kappa ^2}-\frac{5 \left(60 K^2+7\right)}{3456 \sqrt{3} M \kappa }+\frac{\kappa }{8 \sqrt{3} M}-\frac{i K}{3 \sqrt{3} M}\right)\\
&&\displaystyle+L^3 \left(\frac{i K \left(940 K^2-119\right)}{11664 \sqrt{3} M \kappa ^2}+\frac{35 \left(60 K^2+7\right)}{20736 \sqrt{3} M \kappa }-\frac{5 \kappa }{48 \sqrt{3} M}+\frac{i K}{3 \sqrt{3} M}\right)+\mathcal{O}\left(L^4,\frac{1}{\kappa ^3}\right)
\end{array}
\end{equation}

For the Dirac field, we obtain the position of the maximum as follows
\begin{equation}\label{rmax-Dirac}
\begin{array}{rcl}
r_{\max } &=& \displaystyle\frac{11 M}{16 \sqrt{3} \kappa ^3}-\frac{\sqrt{3} M}{2 \kappa }+3 M\\
&&\displaystyle+L \left(\frac{\sqrt{3} M}{4 \kappa }-\frac{11 \sqrt{3} M}{32 \kappa ^3}\right)\\
&&\displaystyle+L^2 \left(\frac{55 \sqrt{3} M}{128 \kappa ^3}-\frac{3 \sqrt{3} M}{16 \kappa }\right)\\
&&\displaystyle+L^3 \left(\frac{5 \sqrt{3} M}{32 \kappa }-\frac{385 M}{256 \sqrt{3} \kappa ^3}\right)+\mathcal{O}\left(L^4,\frac{1}{\kappa ^4}\right)
\end{array}
\end{equation}
and the quasinormal modes are
\begin{equation}\label{eikonal-Dirac}
\begin{array}{rcl}
\omega  &=& \displaystyle\frac{i K \left(119-940 K^2\right)}{46656 \sqrt{3} M \kappa ^2}-\frac{60 K^2+7}{1296 \sqrt{3} M \kappa }+\frac{\kappa }{3 \sqrt{3} M}-\frac{i K}{3 \sqrt{3} M}\\
&&\displaystyle+L \left(\frac{i K \left(940 K^2-119\right)}{23328 \sqrt{3} M \kappa ^2}+\frac{60 K^2+7}{864 \sqrt{3} M \kappa }-\frac{\kappa }{6 \sqrt{3} M}+\frac{i K}{3 \sqrt{3} M}\right)\\
&&\displaystyle+L^2 \left(\frac{i K \left(119-940 K^2\right)}{15552 \sqrt{3} M \kappa ^2}-\frac{5 \left(60 K^2+7\right)}{3456 \sqrt{3} M \kappa }+\frac{\kappa }{8 \sqrt{3} M}-\frac{i K}{3 \sqrt{3} M}\right)\\
&&\displaystyle+L^3 \left(\frac{i K \left(940 K^2-119\right)}{11664 \sqrt{3} M \kappa ^2}+\frac{35 \left(60 K^2+7\right)}{20736 \sqrt{3} M \kappa }-\frac{5 \kappa }{48 \sqrt{3} M}+\frac{i K}{3 \sqrt{3} M}\right)+\mathcal{O}\left(L^4,\frac{1}{\kappa ^3}\right).
\end{array}
\end{equation}


Comparison of the quasinromal frequencies calculated via the obtained analytic formulas are shown in tables IX-XI and comparison with the 6th order WKB formula shows very good agreement for all $\ell$, except $\ell =0$ of the scalar field perturbations.

\section{Conclusions}

Global monopoles hold significance in the realm of physics as they stem from the spontaneous breakdown of symmetries within specific field theories. Their existence offers insights into the structure of the early universe and the evolution of cosmic defects. Delving into the study of global monopoles aids in unraveling the intricate relationship between particle physics and cosmology. This exploration contributes to comprehending the distribution of matter and energy on a large scale, shedding light on the fundamental forces that shaped the cosmos during its formative stages.

In this current work, we calculated quasinormal modes of the scalar and Dirac fields around asymptotically flat bumblebee black holes coupled with a global monopole. We have demonstrated that the quasinormal modes experience significant suppression due to the global monopole parameter, while the influence of the LSB parameter is more gentle. We observe that a form of universal behavior emerges in the vicinity of the extreme limit.

In addition we have derived sufficiently precise analytic approximations for quasinormal modes of the scalar, Dirac and Maxwell fields in the particular limits - when either the bumblebee parameter or the deficit angle are zero. These analytic expressions are obtained via using the higher order WKB method and expansion in terms of the inverse multipole number.

Exploring the Leaver method could facilitate the determination of higher overtones within the spectrum, which characterize the event horizon properties \cite{Konoplya:2022pbc}. Similarly, our analysis can be extended to encompass cases with a non-zero cosmological constant, which holds intrigue due to the emergence of a novel branch of (purely de Sitter) modes. These modes are relevant to the Strong Cosmic Censorship conjecture \cite{Cardoso:2017soq,Hod:2018lmi,Konoplya:2007zx,Dias:2018ynt,Mo:2018nnu,Konoplya:2022kld,Eperon:2019viw,Liu:2019lon}. Furthermore, the investigation of grey-body factors and Hawking radiation associated with the aforementioned black holes remains a pivotal avenue that we leave for future exploration.

\vspace{5mm}
Data Availability Statement: No Data associated in the manuscript

\begin{acknowledgments}
The author would like to thank Roman Konoplya for valuable discussions and help.
\end{acknowledgments}

\bibliographystyle{unsrt}
\bibliography{BibVirtual}

\begin{thebibliography}{10}

\bibitem{Bailey:2006fd}
Quentin~G. Bailey and V.~Alan Kostelecky.
\newblock {Signals for Lorentz violation in post-Newtonian gravity}.
\newblock {\em Phys. Rev. D}, 74:045001, 2006.

\bibitem{Bluhm:2004ep}
Robert Bluhm and V.~Alan Kostelecky.
\newblock {Spontaneous Lorentz violation, Nambu-Goldstone modes, and gravity}.
\newblock {\em Phys. Rev. D}, 71:065008, 2005.

\bibitem{Kostelecky:2010ze}
Alan~V. Kostelecky and Jay~D. Tasson.
\newblock {Matter-gravity couplings and Lorentz violation}.
\newblock {\em Phys. Rev. D}, 83:016013, 2011.

\bibitem{Bluhm:2007bd}
Robert Bluhm, Shu-Hong Fung, and V.~Alan Kostelecky.
\newblock {Spontaneous Lorentz and Diffeomorphism Violation, Massive Modes, and
  Gravity}.
\newblock {\em Phys. Rev. D}, 77:065020, 2008.

\bibitem{Ovgun:2018xys}
Ali \"Ovg\"un, Kimet Jusufi, and \.Izzet Sakall\i{}.
\newblock {Exact traversable wormhole solution in bumblebee gravity}.
\newblock {\em Phys. Rev. D}, 99(2):024042, 2019.

\bibitem{Casana:2017jkc}
R.~Casana, A.~Cavalcante, F.~P. Poulis, and E.~B. Santos.
\newblock {Exact Schwarzschild-like solution in a bumblebee gravity model}.
\newblock {\em Phys. Rev. D}, 97(10):104001, 2018.

\bibitem{Bertolami:2005bh}
O.~Bertolami and J.~Paramos.
\newblock {The Flight of the bumblebee: Vacuum solutions of a gravity model
  with vector-induced spontaneous Lorentz symmetry breaking}.
\newblock {\em Phys. Rev. D}, 72:044001, 2005.

\bibitem{Li:2020dln}
Zonghai Li and Ali \"Ovg\"un.
\newblock {Finite-distance gravitational deflection of massive particles by a
  Kerr-like black hole in the bumblebee gravity model}.
\newblock {\em Phys. Rev. D}, 101(2):024040, 2020.

\bibitem{Ovgun:2018ran}
Ali Ovg\"un, Kimet Jusufi, and Izzet Sakalli.
\newblock {Gravitational lensing under the effect of Weyl and bumblebee
  gravities: Applications of Gauss\textendash{}Bonnet theorem}.
\newblock {\em Annals Phys.}, 399:193--203, 2018.

\bibitem{Li:2020wvn}
Zonghai Li, Guodong Zhang, and Ali \"Ovg\"un.
\newblock {Circular Orbit of a Particle and Weak Gravitational Lensing}.
\newblock {\em Phys. Rev. D}, 101(12):124058, 2020.

\bibitem{Oliveira:2018oha}
R.~Oliveira, D.~M. Dantas, Victor Santos, and C.~A.~S. Almeida.
\newblock {Quasinormal modes of bumblebee wormhole}.
\newblock {\em Class. Quant. Grav.}, 36(10):105013, 2019.

\bibitem{Kanzi:2021cbg}
Sara Kanzi and \.Izzet Sakall\i{}.
\newblock {Greybody radiation and quasinormal modes of Kerr-like black hole in
  Bumblebee gravity model}.
\newblock {\em Eur. Phys. J. C}, 81(6):501, 2021.

\bibitem{Oliveira:2021abg}
R.~Oliveira, D.~M. Dantas, and C.~A.~S. Almeida.
\newblock {Quasinormal frequencies for a black hole in a bumblebee gravity}.
\newblock {\em EPL}, 135(1):10003, 2021.

\bibitem{Kanzi:2022vhp}
Sara Kanzi and \.Izzet Sakall\i{}.
\newblock {Reply to \textquotedblleft{}Comment on \textquoteleft{}Greybody
  radiation and quasinormal modes of Kerr-like black hole in Bumblebee gravity
  model\textquotedblright{}\textquoteright{}}.
\newblock {\em Eur. Phys. J. C}, 82(1):93, 2022.

\bibitem{Maluf:2022knd}
Roberto~V. Maluf and Celio~R. Muniz.
\newblock {Comment on \textquotedblleft{}Greybody radiation and quasinormal
  modes of Kerr-like black hole in Bumblebee gravity
  model\textquotedblright{}}.
\newblock {\em Eur. Phys. J. C}, 82(1):94, 2022.

\bibitem{Gogoi:2022wyv}
Dhruba~Jyoti Gogoi and Umananda~Dev Goswami.
\newblock {Quasinormal modes and Hawking radiation sparsity of GUP corrected
  black holes in bumblebee gravity with topological defects}.
\newblock {\em JCAP}, 06(06):029, 2022.

\bibitem{Uniyal:2022xnq}
Akhil Uniyal, Sara Kanzi, and \.Izzet Sakall\i{}.
\newblock {Some observable physical properties of the higher dimensional dS/AdS
  black holes in Einstein-bumblebee gravity theory}.
\newblock {\em Eur. Phys. J. C}, 83(7):668, 2023.

\bibitem{Liu:2022dcn}
Wentao Liu, Xiongjun Fang, Jiliang Jing, and Jieci Wang.
\newblock {QNMs of slowly rotating Einstein\textendash{}Bumblebee black hole}.
\newblock {\em Eur. Phys. J. C}, 83(1):83, 2023.

\bibitem{Chen:2023cjd}
Chengjia Chen, Qiyuan Pan, and Jiliang Jing.
\newblock {Quasinormal modes of a scalar perturbation around a rotating
  BTZ-like black hole in Einstein-bumblebee gravity}.
\newblock 2 2023.

\bibitem{Mangut:2023oxa}
Mert Mangut, Huriye G\"ursel, Sara Kanzi, and \.Izzet Sakall\i{}.
\newblock {Probing the Lorentz Invariance Violation via Gravitational Lensing
  and Analytical Eigenmodes of Perturbed Slowly Rotating Bumblebee Black
  Holes}.
\newblock {\em Universe}, 9(5):225, 2023.

\bibitem{Filho:2023etf}
A.~A.~Ara\'ujo Filho, H.~Hassanabadi, N.~Heidari, J.~Kr\'\i{}z, P.~J.
  Porf\'\i{}rio, and S.~Zare.
\newblock {Gravitational traces of bumblebee gravity in metric-affine
  formalism}.
\newblock 5 2023.

\bibitem{Konoplya:2006rv}
R.~A. Konoplya and A.~Zhidenko.
\newblock {Perturbations and quasi-normal modes of black holes in
  Einstein-Aether theory}.
\newblock {\em Phys. Lett. B}, 644:186--191, 2007.

\bibitem{Konoplya:2006ar}
R.~A. Konoplya and A.~Zhidenko.
\newblock {Gravitational spectrum of black holes in the Einstein-Aether
  theory}.
\newblock {\em Phys. Lett. B}, 648:236--239, 2007.

\bibitem{Ding:2017gfw}
Chikun Ding.
\newblock {Quasinormal ringing of black holes in Einstein-aether theory}.
\newblock {\em Phys. Rev. D}, 96(10):104021, 2017.

\bibitem{Ding:2018mmh}
Chi-Kun Ding.
\newblock {Gravitational Perturbations in Einstein Aether Black Hole
  Spacetime}.
\newblock {\em Chin. Phys. Lett.}, 35(10):100401, 2018.

\bibitem{Xiong:2021cth}
Wei Xiong, Peng Liu, Cheng-Yong Zhang, and Chao Niu.
\newblock {Quasinormal modes of the Einstein-Maxwell-aether black hole}.
\newblock {\em Phys. Rev. D}, 106(6):064057, 2022.

\bibitem{Kokkotas:1999bd}
Kostas~D. Kokkotas and Bernd~G. Schmidt.
\newblock {Quasinormal modes of stars and black holes}.
\newblock {\em Living Rev. Rel.}, 2:2, 1999.

\bibitem{Nollert:1999ji}
Hans-Peter Nollert.
\newblock {TOPICAL REVIEW: Quasinormal modes: the characteristic `sound' of
  black holes and neutron stars}.
\newblock {\em Class. Quant. Grav.}, 16:R159--R216, 1999.

\bibitem{Konoplya:2011qq}
R.~A. Konoplya and A.~Zhidenko.
\newblock {Quasinormal modes of black holes: From astrophysics to string
  theory}.
\newblock {\em Rev. Mod. Phys.}, 83:793--836, 2011.

\bibitem{Barack:2018yly}
Leor Barack et~al.
\newblock {Black holes, gravitational waves and fundamental physics: a
  roadmap}.
\newblock {\em Class. Quant. Grav.}, 36(14):143001, 2019.

\bibitem{EventHorizonTelescope:2022xqj}
Kazunori Akiyama et~al.
\newblock {First Sagittarius A* Event Horizon Telescope Results. VI. Testing
  the Black Hole Metric}.
\newblock {\em Astrophys. J. Lett.}, 930(2):L17, 2022.

\bibitem{Goddi:2016qax}
C.~Goddi et~al.
\newblock {BlackHoleCam: Fundamental physics of the galactic center}.
\newblock {\em Int. J. Mod. Phys. D}, 26(02):1730001, 2016.

\bibitem{Vilenkin1981}
Alexander Vilenkin.
\newblock Gravitational field of vacuum domain walls and strings.
\newblock {\em Physical Review D}, 23(4):852, 1981.

\bibitem{Lin:2023foj}
Rui-Hui Lin, Rui Jiang, and Xiang-Hua Zhai.
\newblock {Quasinormal modes of the spherical bumblebee black holes with a
  global monopole}.
\newblock {\em Eur. Phys. J. C}, 83(8):720, 2023.

\bibitem{Vilenkin:1981kz}
A.~Vilenkin.
\newblock {Cosmic Strings}.
\newblock {\em Phys. Rev. D}, 24:2082--2089, 1981.

\bibitem{Barriola:1989hx}
Manuel Barriola and Alexander Vilenkin.
\newblock {Gravitational Field of a Global Monopole}.
\newblock {\em Phys. Rev. Lett.}, 63:341, 1989.

\bibitem{Gullu:2020qzu}
\.Ibrahim G\"ull\"u and Ali \"Ovg\"un.
\newblock {Schwarzschild-like black hole with a topological defect in bumblebee
  gravity}.
\newblock {\em Annals Phys.}, 436:168721, 2022.

\bibitem{Abdalla:2006qj}
E.~Abdalla, B.~Cuadros-Melgar, A.~B. Pavan, and C.~Molina.
\newblock {Stability and thermodynamics of brane black holes}.
\newblock {\em Nucl. Phys. B}, 752:40--59, 2006.

\bibitem{Konoplya:2022iyn}
R.~A. Konoplya.
\newblock {Quasinormal modes in higher-derivative gravity: Testing the black
  hole parametrization and sensitivity of overtones}.
\newblock {\em Phys. Rev. D}, 107(6):064039, 2023.

\bibitem{Zhidenko:2003wq}
A.~Zhidenko.
\newblock {Quasinormal modes of Schwarzschild de Sitter black holes}.
\newblock {\em Class. Quant. Grav.}, 21:273--280, 2004.

\bibitem{Schutz:1985km}
Bernard~F. Schutz and Clifford~M. Will.
\newblock {Black hole normal modes: a semianalytic approach}.
\newblock {\em Astrophys. J. Lett.}, 291:L33--L36, 1985.

\bibitem{Iyer:1986np}
Sai Iyer and Clifford~M. Will.
\newblock {Black Hole Normal Modes: A {WKB} Approach. 1. Foundations and
  Application of a Higher Order {WKB} Analysis of Potential Barrier
  Scattering}.
\newblock {\em Phys. Rev. D}, 35:3621, 1987.

\bibitem{Konoplya:2003ii}
R.~A. Konoplya.
\newblock {Quasinormal behavior of the d-dimensional Schwarzschild black hole
  and higher order WKB approach}.
\newblock {\em Phys. Rev. D}, 68:024018, 2003.

\bibitem{Matyjasek:2017psv}
Jerzy Matyjasek and Micha\l{} Opala.
\newblock {Quasinormal modes of black holes. The improved semianalytic
  approach}.
\newblock {\em Phys. Rev. D}, 96(2):024011, 2017.

\bibitem{Konoplya:2019hlu}
R.~A. Konoplya, A.~Zhidenko, and A.~F. Zinhailo.
\newblock {Higher order WKB formula for quasinormal modes and grey-body
  factors: recipes for quick and accurate calculations}.
\newblock {\em Class. Quant. Grav.}, 36:155002, 2019.

\bibitem{Konoplya:2023ahd}
R.~A. Konoplya, D.~Ovchinnikov, and B.~Ahmedov.
\newblock {Bardeen spacetime as a quantum corrected Schwarzschild black hole:
  Quasinormal modes and Hawking radiation, e-Print:2307.10801 (2023)}.

\bibitem{Konoplya:2023ppx}
R.~A. Konoplya.
\newblock {Quasinormal modes and grey-body factors of regular black holes with
  a scalar hair from the Effective Field Theory}.
\newblock {\em JCAP}, 07:001, 2023.

\bibitem{Matyjasek:2021xfg}
Jerzy Matyjasek.
\newblock {Accurate quasinormal modes of the five-dimensional
  Schwarzschild-Tangherlini black holes}.
\newblock {\em Phys. Rev. D}, 104(8):084066, 2021.

\bibitem{Guo:2023ivz}
Guangzhou Guo, Peng Wang, Houwen Wu, and Haitang Yang.
\newblock {Superradiance instabilities of charged black holes in
  Einstein-Maxwell-scalar theory}.
\newblock {\em JHEP}, 07:070, 2023.

\bibitem{Chrysostomou:2022evl}
Anna Chrysostomou, Alan Cornell, Aldo Deandrea, \'Etienne Ligout, and Dimitrios
  Tsimpis.
\newblock {Black holes and nilmanifolds: quasinormal modes as the fingerprints
  of extra dimensions?}
\newblock {\em Eur. Phys. J. C}, 83(4):325, 2023.

\bibitem{Konoplya:2019xmn}
R.~A. Konoplya.
\newblock {Quantum corrected black holes: quasinormal modes, scattering,
  shadows}.
\newblock {\em Phys. Lett. B}, 804:135363, 2020.

\bibitem{Konoplya:2019ppy}
R.~A. Konoplya and A.~F. Zinhailo.
\newblock {Hawking radiation of non-Schwarzschild black holes in higher
  derivative gravity: a crucial role of grey-body factors}.
\newblock {\em Phys. Rev. D}, 99(10):104060, 2019.

\bibitem{Konoplya:2010vz}
R.~A. Konoplya and A.~Zhidenko.
\newblock {Long life of Gauss-Bonnet corrected black holes}.
\newblock {\em Phys. Rev. D}, 82:084003, 2010.

\bibitem{Matyjasek:2020bzc}
Jerzy Matyjasek.
\newblock {Quasinormal modes of dirty black holes in the effective theory of
  gravity with a third order curvature term}.
\newblock {\em Phys. Rev. D}, 102(12):124046, 2020.

\bibitem{Konoplya:2006gq}
R.~A. Konoplya, A.~Zhidenko, and C.~Molina.
\newblock {Late time tails of the massive vector field in a black hole
  background}.
\newblock {\em Phys. Rev. D}, 75:084004, 2007.

\bibitem{Gundlach:1993tp}
Carsten Gundlach, Richard~H. Price, and Jorge Pullin.
\newblock {Late time behavior of stellar collapse and explosions: 1. Linearized
  perturbations}.
\newblock {\em Phys. Rev. D}, 49:883--889, 1994.

\bibitem{Bronnikov:2019sbx}
Kirill~A. Bronnikov and Roman~A. Konoplya.
\newblock {Echoes in brane worlds: ringing at a black hole--wormhole
  transition}.
\newblock {\em Phys. Rev. D}, 101(6):064004, 2020.

\bibitem{Konoplya:2020jgt}
R.~A. Konoplya, A.~F. Zinhailo, and Z.~Stuchlik.
\newblock {Quasinormal modes and Hawking radiation of black holes in cubic
  gravity}.
\newblock {\em Phys. Rev. D}, 102(4):044023, 2020.

\bibitem{Leaver:1985ax}
E.~W. Leaver.
\newblock {An Analytic representation for the quasi normal modes of Kerr black
  holes}.
\newblock {\em Proc. Roy. Soc. Lond. A}, 402:285--298, 1985.

\bibitem{Cardoso:2008bp}
Vitor Cardoso, Alex~S. Miranda, Emanuele Berti, Helvi Witek, and Vilson~T.
  Zanchin.
\newblock {Geodesic stability, Lyapunov exponents and quasinormal modes}.
\newblock {\em Phys. Rev. D}, 79(6):064016, 2009.

\bibitem{Jusufi:2019ltj}
Kimet Jusufi.
\newblock {Quasinormal Modes of Black Holes Surrounded by Dark Matter and Their
  Connection with the Shadow Radius}.
\newblock {\em Phys. Rev. D}, 101(8):084055, 2020.

\bibitem{Jafarzade:2020ova}
Khadije Jafarzade, Mahdi Kord~Zangeneh, and Francisco S.~N. Lobo.
\newblock {Shadow, deflection angle and quasinormal modes of Born-Infeld
  charged black holes}.
\newblock {\em JCAP}, 04:008, 2021.

\bibitem{Konoplya:2017wot}
R.~A. Konoplya and Z.~Stuchl\'\i{}k.
\newblock {Are eikonal quasinormal modes linked to the unstable circular null
  geodesics?}
\newblock {\em Phys. Lett. B}, 771:597--602, 2017.

\bibitem{Konoplya:2022gjp}
R.~A. Konoplya.
\newblock {Further clarification on quasinormal modes/circular null geodesics
  correspondence}.
\newblock {\em Phys. Lett. B}, 838:137674, 2023.

\bibitem{Konoplya:2023moy}
R.~A. Konoplya and A.~Zhidenko.
\newblock {Analytic expressions for quasinormal modes and grey-body factors in
  the eikonal limit and beyond}.
\newblock {\em Class. Quant. Grav.}, 40(24):245005, 2023.

\bibitem{Malik:2024sxv}
Zainab Malik.
\newblock {Quasinormal Modes of Dilaton Black Holes: Analytic Approximations}.
\newblock {\em Int. J. Theor. Phys.}, 63(5):128, 2024.

\bibitem{Malik:2024voy}
Zainab Malik.
\newblock {Analytic expressions for quasinormal modes of the
  Reissner\textendash{}Nordstr\"om-like black holes}.
\newblock {\em Int. J. Mod. Phys. A}, 39(05n06):2450024, 2024.

\bibitem{Konoplya:2022pbc}
R.~A. Konoplya and A.~Zhidenko.
\newblock {First few overtones probe the event horizon geometry}.
\newblock 9 2022.

\bibitem{Cardoso:2017soq}
Vitor Cardoso, Jo\~ao~L. Costa, Kyriakos Destounis, Peter Hintz, and Aron
  Jansen.
\newblock {Quasinormal modes and Strong Cosmic Censorship}.
\newblock {\em Phys. Rev. Lett.}, 120(3):031103, 2018.

\bibitem{Hod:2018lmi}
Shahar Hod.
\newblock {Quasinormal modes and strong cosmic censorship in near-extremal
  Kerr\textendash{}Newman\textendash{}de Sitter black-hole spacetimes}.
\newblock {\em Phys. Lett. B}, 780:221--226, 2018.

\bibitem{Konoplya:2007zx}
R.~A. Konoplya and A.~Zhidenko.
\newblock {Decay of a charged scalar and Dirac fields in the Kerr-Newman-de
  Sitter background}.
\newblock {\em Phys. Rev. D}, 76(8):084018, 2007.
\newblock [Erratum: Phys.Rev.D 90, 029901 (2014)].

\bibitem{Dias:2018ynt}
Oscar J.~C. Dias, Felicity~C. Eperon, Harvey~S. Reall, and Jorge~E. Santos.
\newblock {Strong cosmic censorship in de Sitter space}.
\newblock {\em Phys. Rev. D}, 97(10):104060, 2018.

\bibitem{Mo:2018nnu}
Yuyu Mo, Yu~Tian, Bin Wang, Hongbao Zhang, and Zhen Zhong.
\newblock {Strong cosmic censorship for the massless charged scalar field in
  the Reissner-Nordstrom\textendash{}de Sitter spacetime}.
\newblock {\em Phys. Rev. D}, 98(12):124025, 2018.

\bibitem{Konoplya:2022kld}
R.~A. Konoplya and A.~Zhidenko.
\newblock {How general is the strong cosmic censorship bound for quasinormal
  modes?}
\newblock {\em JCAP}, 11:028, 2022.

\bibitem{Eperon:2019viw}
Felicity~C. Eperon, Bogdan Ganchev, and Jorge~E. Santos.
\newblock {Plausible scenario for a generic violation of the weak cosmic
  censorship conjecture in asymptotically flat four dimensions}.
\newblock {\em Phys. Rev. D}, 101(4):041502, 2020.

\bibitem{Liu:2019lon}
Hang Liu, Ziyu Tang, Kyriakos Destounis, Bin Wang, Eleftherios Papantonopoulos,
  and Hongbao Zhang.
\newblock {Strong Cosmic Censorship in higher-dimensional
  Reissner-Nordstr\"om-de Sitter spacetime}.
\newblock {\em JHEP}, 03:187, 2019.

\end{thebibliography}
\end{document}